\documentclass[aps,prb,preprint,groupedaddress]{revtex4}

\usepackage{graphicx}
\usepackage{bm}

\begin{document}

\title{Electron scattering from a mesoscopic disk in Rashba system}

\author{Jr-Yu Yeh}
\affiliation{Department of Physics, National Taiwan Normal
University, Taipei, Taiwan}
\author{Chung-Yu Mou}
\affiliation{Department of Physics, National Tsing Hwa University,
Hsinchu, Taiwan}
\author{Ming-Che Chang}
\email[corresponding e-mail address:]{ changmc@phy.ntnu.edu.tw}
\affiliation{Department of Physics, National Taiwan Normal
University, Taipei, Taiwan}

\date{\today}
\begin{abstract}
Electrons with spin-orbit coupling moving in mesoscopic structures
can often exhibit local spin polarization. In this paper, we study
the influence of the Rashba coupling on the scattering of
two-dimensional electrons from a circular disk. It is observed
that spin-polarized regions exist, even if the incident electrons
are unpolarized. In addition to the distributions of charge and
spin current in the near-field region, we also analyze the
symmetry and the differential cross-section of the scattering.
\end{abstract}

\pacs{ 72.25.-b; 
72.25.Dc; 
73.40.-c 
}

\maketitle

\section{Introduction}

Spin-orbit interaction influences the electronic and transport
properties of semiconductors. For example, it lifts the degeneracy
of the valence bands, modifies the electron g-factor,\cite{kittel}
and causes skew scattering in the presence of (spinless)
impurities. Such a skew scattering is a possible mechanism for the
extrinsic spin Hall effect.\cite{hirsch} In addition, spin-orbit
interaction plays an important role in the recently proposed
intrinsic spin Hall effect.\cite{murakami,sinova} It is also
crucial in the mechanisms of spin relaxation and optical
orientation in semiconductors.\cite{zutic04,meier}

It is highly desirable to generate flows of polarized spins in
semiconductors with the help of the spin-orbit interaction. In
these endeavors, the Rashba spin-orbit coupling\cite{bychkov84} in
two-dimensional electron gas (2DEG) plays a special role since it
allows manipulation of spin flows by varying the gate bias. This
has motivated several creative proposals for its
application.\cite{datta90} To explore the possibilities, the
effect of Rashba coupling in many types of mesoscopic structure
have been investigated, such as a quantum wire,\cite{wire} a
quantum ring,\cite{ring} and a quantum dot.\cite{dot,negative} In
several studies, it was found that a device with a simple
geometry, in combination with the Rashba coupling, could serve as
a spin filter. For example, the device could be a T-shaped
channel,\cite{T} a quantum point contact,\cite{QPC} parallel
interfaces that cause double refraction,\cite{double_refraction}
or even just a curved wire.\cite{curve} After applying a magnetic
field, we could further build a spin filter based on electron
focusing,\cite{focusing} or based on the an interferometer of the
Stern-Gerlach type.\cite{SG}

In this paper, we study the scattering of electrons by a disk in
the 2DEG with Rashba coupling.\cite{lapidus} The analysis can be
applied to a wide range of situations in which the radius $R$ of
the disk can be much smaller, roughly the same, or much larger
than the electron wave length $\lambda$. For example, the usual
impurity scattering can be simulated with $R\ll\lambda$, while the
scattering by an artificial mesoscopic disk corresponds to
$R\approx\lambda$. Here we focus on the latter case to search for
possible effect of local spin generation. We find that, because of
the Rashba coupling, spin-up and spin-down electrons are indeed
separated and accumulate in regions of curved stripes. The
associated charge and spin currents are analyzed in details. In
addition, the properties of symmetry, as well as the differential
cross-section, are also investigated.\cite{thesis}

This paper is organized as follows: Sec. II is the theoretical
analysis. In Sec. III, the results from numerical calculations are
presented. Sec. IV is the conclusion.

\section{Analysis of the disk scattering}

\subsection{Hard disk scattering}

Consider a two-dimensional electron system with a circular disk at
the origin,
\begin{equation}
H=\frac{p^2}{2m}+\frac{\alpha}{\hbar}(\sigma_x p_y-\sigma_y
p_x)+V(r),
\end{equation}
where $V(r)=V_0$ when $r\leq R$ and 0 otherwise. In the following,
the potential of the disk is considered infinite (``hard" disk).
The two-component wave function $(\psi_1,\psi_2)^T$ with energy
$E$ satisfies the coupled equations:
\begin{eqnarray}
    \left(\frac{\partial^2}{\partial
    r^2}+\frac{1}{r}\frac{\partial}{\partial r}
    +\frac{1}{r^2}\frac{\partial^2}{\partial \phi^2}+\frac{2mE}{\hbar^2} \right
    )\psi_1
    +\frac{2im\alpha}{\hbar^2}e^{-i\phi}\left(i\frac{\partial}{\partial
    r}+\frac{1}{r}\frac{\partial}{\partial \phi}\right )\psi_2 &=&0 \nonumber \\
    \left(\frac{\partial^2}{\partial
    r^2}+\frac{1}{r}\frac{\partial}{\partial r}
    +\frac{1}{r^2}\frac{\partial^2}{\partial \phi^2}+\frac{2mE}{\hbar^2} \right
    )\psi_2
    +\frac{2im\alpha}{\hbar^2}e^{i\phi}\left(-i\frac{\partial}{\partial
    r} +\frac{1}{r}\frac{\partial}{\partial \phi}\right )\psi_1 &=&0.
\end{eqnarray}
It follows that the energy eigenstates of the Hamiltonian with
momentum $k$, helicity $\eta$ ($E=\hbar^2k^2/2m+\eta\alpha k,
\eta=\pm$), and angular momentum $(n+1/2)\hbar$ (where $n$ is an
integer) are
\begin{equation}
\psi_{\eta n}(r,\phi)=\left(
    \begin{array}{c}
    \Omega_{\eta n}(kr)e^{in\phi}\\
    \Omega_{\eta(n+1)}(kr)e^{i(n+1)\phi}
    \end{array}\right),
\end{equation}
where $\Omega_{\eta n}$ can be a Bessel function, a Neumann
function, or their linear combination, such as a Hankel function.
We will choose the Hankel functions as the eigen-basis since their
behavior at large radius suits the boundary condition for
scattering.

The energy eigenstate of the Schroedinger equation with energy $E$
and angular momentum $(n+1/2)\hbar$ can be written as
\begin{eqnarray}
\Psi_n&=&a_n\left(
    \begin{array}{c}
    H^1_n(kr)e^{in\phi}\\H^1_{n+1}(kr)e^{i(n+1)\phi}
    \end{array}\right )
    +b_n\left(
    \begin{array}{c}
    H^2_n(kr)e^{in\phi}\\H^2_{n+1}(kr)e^{i(n+1)\phi}
    \end{array}\right )\nonumber \\
    &+&c_n\left(
    \begin{array}{c}
    H^1_n(k'r)e^{in\phi}\\-H^1_{n+1}(k'r)e^{i(n+1)\phi}
    \end{array}\right )
    +d_n\left(
    \begin{array}{c}
    H^2_n(k'r)e^{in\phi}\\-H^2_{n+1}(k'r)e^{i(n+1)\phi}
    \end{array}\right ),
\end{eqnarray}
where $E=\hbar^2k^2/2m+\alpha k=\hbar^2{k'}^2/2m-\alpha{k'}$. The
first two terms have positive helicity; while the other two terms
have negative helicity. The most general eigenstate of the
Schroedinger equation with energy $E$ is a superposition of the
$\Psi_n$'s, where the coefficients $a_n, b_n, c_n$, and $d_n$ are
determined by boundary conditions. Because of the circular
symmetry of the potential, the angular momentum is conserved for
each $n$-component. Therefore, each component can be considered
independent during the scattering.

The incident plane wave with momentum $k$ and helicity $\eta$ can
be decomposed as the following linear superposition:
\begin{equation}
\Psi_{{\rm in},\eta}=e^{ikx}\frac{1}{\sqrt{2}}\left(
    \begin{array}{c}
    1\\
    -\eta i
    \end{array}\right )
    =\frac{1}{\sqrt{2}}\sum^\infty_{n=-\infty}\left(
    \begin{array}{c}
    i^nJ_n(kr)e^{in\phi}\\
    \eta i^nJ_{n+1}(kr)e^{i(n+1)\phi}
    \end{array}\right )
\end{equation}
The Bessel functions $J_n(kr)$ can be further decomposed as Hankel
functions $H_n^1(kr)$ and $H_n^2(kr)$, the former correspond to
outgoing circular waves, while the later correspond to incoming
circular waves (with no phase shift). If an incident wave has a
definite helicity, $\Psi_{{\rm in},+}$, then by comparing Eq.~(4)
with the components of Eq.~(5) at large distance, one will obtain
$b_n=i^n/2\sqrt{2}$ and $d_n=0$ (i.e. no incoming circular wave
with negative helicity) for all $n$. The coefficients $a_n$ and
$c_n$ need to be determined from the boundary condition at $r=R$.
For a hard disk, it can be shown that
\begin{eqnarray}
    a_n &=& -\frac{i^n}{2\sqrt{2}}\frac{H_n^1(\tilde k')H_{n+1}^2(\tilde k)
    +H_n^2(\tilde k)H_{n+1}^1(\tilde k')}{H_n^1(\tilde k')H_{n+1}^1(\tilde k)
    +H_n^1(\tilde k)H_{n+1}^1(\tilde k')},\nonumber \\
    c_n &=&\frac{i^n}{2\sqrt{2}}\frac{H_n^1(\tilde k)H_{n+1}^2(\tilde k)
    -H_n^2(\tilde k)H_{n+1}^1(\tilde k)}{H_n^1(\tilde k')H_{n+1}^1(\tilde k)
    +H_n^1(\tilde k)H_{n+1}^1(\tilde k')},
    \label{ac}
\end{eqnarray}
where ${\tilde k}\equiv kR$ and ${\tilde k}'\equiv k'R$. Notice
that the nonzero probability amplitudes $c_n$ lead to outgoing
waves with flipped helicity. For reference, if the incident wave
is $\Psi_{{\rm in},-}$, then $b_n=0$ and $d_n=i^n/2\sqrt{2}$. At
the mean time, the roles of $k$ and $k'$, as well as the roles of
$a_n$ and $c_n$, have to be interchanged.

A note on the unitary condition: for convenience of discussion,
consider an incoming wave with positive helicity and angular
momentum $(n+1/2)\hbar$ (i.e., the $b_n$-wavelet). Because the
angular momentum is conserved during the scattering, the electron
can only be scattered to $a_n$ and $c_n$ channels with the same
$n$. From particle conservation at large distance, one expects
that the probability amplitudes in Eq.~(4) should satisfy
$|a_n|^2+(k/k')|c_n|^2=|b_n|^2(=1/8)$ for all $n$, which has
indeed been confirmed in our numerical calculation.

\subsection{Properties of symmetry}

The system has a mirror symmetry with respect to the $x$-axis.
Therefore, by analyzing the Schrodinger equation with $y$ replaced
by $-y$ , one finds $\Psi(\vec{r}^*)=-\sigma_y \Psi(\vec{r})$,
where $\vec{r}=(x,y)$, and $\vec{r}^*\equiv(x,-y)$ is the
mirror-reflected point of $\vec{r}$. Such as relation can also be
obtained by a space inversion of the (three-dimensional)
coordinate, followed by a rotation with respect to the new
$y$-axis by 180 degrees. Consequently, for the expectation value
of the spin, we have
\begin{equation}
(S_x(\vec{r}^*),S_y(\vec{r}^*),S_z(\vec{r}^*))=(-S_x(\vec{r}),
S_y(\vec{r}),-S_z(\vec{r})). \label{symms}
\end{equation}
In a Rashba system, the current density operator is defined as,
\begin{equation}
\vec{j}=\frac{\hbar}{2mi}\left(\Psi^\dagger\frac{d\Psi}
{d\vec{r}}-\frac{d\Psi^\dagger}{d\vec{r}}\Psi\right)
-\frac{\alpha}{\hbar}\Psi^\dagger\vec{\sigma}\times\hat{z}\Psi.
\end{equation}
Therefore, the distribution of the expectation value of the
current density has the following symmetry:
\begin{equation}
(j_x(\bar{r}),j_y(\bar{r}))=(j_x(\bar{r}),-j_y(\bar{r})).
\end{equation}
We adopt the generally accepted definition of the spin current
density operator,\cite{sc} $\vec{j}_s^\gamma={\rm
Re}\Psi^\dagger(\sigma^\gamma \dot{\vec{r}})\Psi$, whose
expectation values have the symmetries,
\begin{eqnarray}
(j^x_x(\vec{r}^*),j^x_y(\vec{r}^*))&=&(-j^x_x(\vec{r}),j^x_y(\vec{r}))\nonumber\\
(j^y_x(\vec{r}^*),j^y_y(\vec{r}^*))&=&(j^y_x(\vec{r}),-j^y_y(\vec{r}))\nonumber\\
(j^z_x(\vec{r}^*),j^z_y(\vec{r}^*))&=&(-j^z_x(\vec{r}),j^z_y(\vec{r})).
\end{eqnarray}
These symmetries will be confirmed by the numerical results in
Sec. III.

\subsection{Asymptotic behavior of the scattered wave}

For convenience, the wave function $\sum\Psi_n$ can be seperated
into an incident plane wave and a scattered wave. At large
distance with $kr\gg 1$, the scattered wave has the asymptotic
form,
\begin{equation}
\Psi_{{\rm sc}}=\frac{e^{ikr}}{\sqrt{r}} \left(\begin{array}{c}
f_1(\phi)\\f_2(\phi)
\end{array}\right )
+\frac{e^{ik'r}}{\sqrt{r}} \left(\begin{array}{c}
g_1(\phi)\\g_2(\phi)
\end{array}\right ),
\label{scattered}
\end{equation}
where
\begin{eqnarray}
    {\bf f}\equiv\left(\begin{array}{c} f_1(\phi)\\f_2(\phi)
    \end{array}\right )
    &=& \sqrt{\frac{2}{\pi k}}\sum_n \left(a_n-b_n\right )e^{-i(n+1/2)\pi/2}
    \left(\begin{array}{c} e^{in\phi}\\-ie^{i(n+1)\phi}
    \end{array}\right ),
    \nonumber \\
    {\bf g}\equiv\left(\begin{array}{c} g_1(\phi)\\g_2(\phi)
    \end{array}\right )
    &=& \sqrt{\frac{2}{\pi k'}}\sum_n \left(c_n-d_n\right )e^{-i(n+1/2)\pi/2}
    \left(\begin{array}{c} e^{in\phi}\\ie^{i(n+1)\phi}
    \end{array}\right ).
    \label{fg}
\end{eqnarray}
It can be shown that ${\bf f}^\dagger\cdot{\bf g}=0$. Also, for
the incoming plane wave ${\Psi}_{{\rm in},+}$,
\begin{eqnarray}
    \sum_{i=1}^2f_i^*\vec{\sigma}f_i&=&|{\bf f}|^2
    (\sin\phi,-\cos\phi,0), \nonumber \\
    \sum_{i=1}^2g_i^*\vec{\sigma}g_i&=&|{\bf g}|^2
    (-\sin\phi,\cos\phi,0).\label{far}
\end{eqnarray}
Therefore, ${\bf f}$-spinor and ${\bf g}$-spinor possess spins
with opposite directions at large distance. Both spins lie on the
plane and are perpendicular to the direction of propagation. One
can obtain the same equations for the incoming wave ${\Psi}_{{\rm
in},-}$, but the signs of the spin expectation values are
opposite.

After a straightforward calculation, one can show that the
scattered current density at large distance is
\begin{equation}
\vec{j}_{{\rm sc}}=\frac{1}{r}\left(\frac{\hbar k}{m}
+\frac{\alpha}{\hbar}\right ) |{\bf
f}|^2\hat{r}+\frac{1}{r}\left(\frac{\hbar
k'}{m}-\frac{\alpha}{\hbar}\right ) |{\bf g}|^2\hat{r},
\end{equation}
from which the differential cross-section $\sigma'(\phi)\equiv
r|\vec{j}_{{\rm sc}}|/|\vec{j}_{{\rm in}}|$ can be calculated. For
incoming waves $\Psi_{{\rm in},+}$ and $\Psi_{{\rm in},-}$, the
current densities $|\vec{j}_{{\rm in},+}|$ and $|\vec{j}_{{\rm
in},-}|$ are $\hbar k/m+\alpha/\hbar$ and $\hbar
k'/m-\alpha/\hbar$ respectively. In fact, they are equal in
magnitude if the two incident waves have the same energy.
Therefore, the differential cross-sections $\sigma_\eta$ for
incoming waves with helicity $\eta$ are
\begin{eqnarray}
\sigma_+'&=&\left|{\bf f}_+\right|^2+\left|{\bf
g}_+\right|^2 \equiv\sigma'_{++}+\sigma'_{+-},\nonumber\\
\sigma_-'&=&\left|{\bf f}_-\right|^2+\left|{\bf g}_-\right|^2
\equiv\sigma'_{-+}+\sigma'_{--} \label{cross-section}
\end{eqnarray}
where ${\bf f}_\eta$ is the ${\bf f}$-spinor in Eq.~(12), but with
the coefficients $a_n$ and $b_n$ suitably chosen for the
scattering of $\Psi_{{\rm in},\eta}$, similarly for ${\bf g}_\eta$
(see the discussion following Eq.~(6)). The differential
cross-sections $\sigma'_{\eta,\eta}$ and $\sigma'_{\eta,-\eta}$
represent helicity-preserved and helicity-flipped scatterings
respectively. If the incoming wave is an incoherent mixture of
both helicities with fractional populations $P_\eta$, then the
differential cross-section is simply the weighted average of the
two differential cross-sections:
$\sigma'=P_+\sigma'_++P_-\sigma'_-$.

\begin{figure}
\includegraphics[width=2.8in]{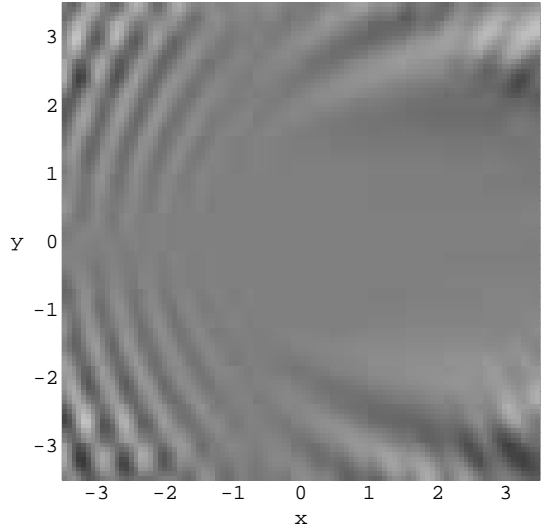}
\caption{Distribution of out-of-plane spin $S_z$, with
$\alpha=(1/10)\hbar^2k/2m$. The incident wave is an equal mixture
of both helicities $\eta=\pm$. Lighter (darker) regions represent
the regions of spin up (down). The peak intensity in this figure
is approximately 0.22 (1 for fully polarized).} \label{fig3}
\end{figure}

\section{Distributions of charge and spin}

In the following, we report on the distributions of spin density,
charge current density, and spin current density, assuming the
wave length of the incident wave $\lambda=R$. We have also studied
the cases with a larger $\lambda$ (e.g. $\lambda=3R$) and a
smaller $\lambda$ (e.g. $\lambda=R/3$). These results are not
presented since the main difference is the change of scales. In
the limit of $\lambda\gg R$, whose scale is more relevant to the
case of impurity scattering, only the components with the smallest
angular momentum ($n$=0 and $-1$) need to be considered. From
Eq.~(\ref{ac}), one finds that both $|c_0|$ and
$|c_{-1}|\rightarrow 0$  as $kR\ll 1$. Therefore, there would be
little change of helicity in the long wave length limit.

For comparison with realistic values, we choose $m=0.068 m_e$ for
electrons in the GaAs-AlGaAs heterojunctions. The radius of the
disk is fixed at $R$=1000 $\rm \AA$. The corresponding Fermi
energy and electron density for $\lambda_F=1000\rm \AA$ are 2.2
meV and $6.3\times 10^{10}$/cm${}^2$ respectively, which are
typical values. To enhance the visual effect of the spin-orbit
coupling, the Rashba energy $\alpha k$ is chosen to be one-tenth
of the kinetic energy $\hbar^2 k^2/2m$, which requires
$\alpha=0.35$ nm-eV, about one order of magnitude larger than the
value in GaAs.

\begin{figure}
\includegraphics[width=2.8in]{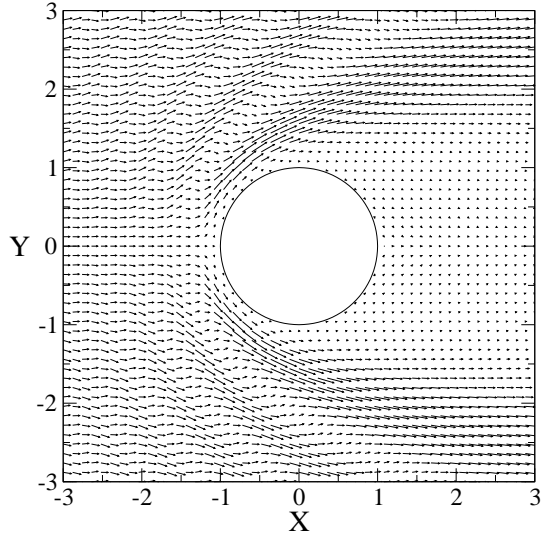}
\caption{Distribution of charge current density near the disk. The
incident wave is $\Psi_{{\rm in},+}$ with wavelength $\lambda=R$.
[$\alpha=(1/10)\hbar^2k/2m$]} \label{fig4}
\end{figure}

The out-of-plane ($z$) spin component results from a
spin-unpolarized incident wave with an equal (incoherent) mixture
of both helicities is plotted in Fig.~1, which is antisymmetric
with respect to the horizontal $x$-axis:
$S_z(\vec{r}^*)=-S_z(\vec{r})$.  Notice that the incident waves
with opposite helicities (but the same energy) have different wave
vectors ($k'-k=2m\alpha/\hbar^2$). Therefore, their interference
patterns for opposite helicities with spins point at opposite
directions are slightly displaced with respect to each other.
Because of such a displacement between $\eta=+$ and $\eta=-$,
regions with net $z$-spin still exist after partial cancellation.
The existence of $S_z$ relies on the scattered part of the wave
function in the near-field region [see Eqs.~(\ref{scattered}) and
(\ref{far})] and would decay to zero at large distance.

Unlike the spins in Fig.~1, the distribution of charge current
(Fig.~2), as well as the differential cross-section (Fig.~3),
which is defined using the {\it charge} current densities, are not
sensitive to the strength of the Rashba coupling, and look very
similar even if the coupling is turned off. Notice that Rashba
spin-orbit coupling in fact preserves the helicity. The major
cause of the helicity flip is the potential $V(r)$, which is
incompatible with the helicity operator. In Fig.~2, the
distribution of charge current density from the scattering of
$\Psi_{{\rm in},+}$ shows the expected pattern of the flow around
the disk. If the incident wave is a mixed state, then instead of
cancellation, the slightly displaced current densities from both
helicities will add up.

\begin{figure}
\includegraphics[width=2.8in]{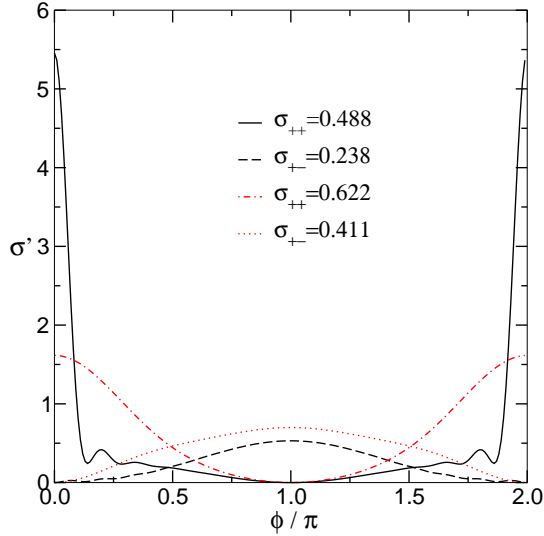}
\caption{(Color online) Differential cross-sections (in units of
$R$) for wave length $\lambda=R$ (solid line for $\sigma'_{++}$,
dashed line for $\sigma'_{+-}$) and $\lambda=10R$ (dash-dotted
line for $\sigma'_{++}$, dotted line for $\sigma'_{+-}$). The
numbers in the legend are total cross-sections after integration
over angle. [$\alpha=(1/10)\hbar^2k/2m$]}\label{fig5}
\end{figure}

The differential cross-sections for $\Psi_{{\rm in},+}$ are shown
in Fig.~3. It can be seen that the helicity-preserved scattering
($\sigma'_{++}$) peaks at the forward direction ($\phi=0, 2\pi$),
while the helicity-flipped scattering ($\sigma'_{+-}$) peaks at
the backward direction ($\phi=\pi$). At the backward direction,
$\sigma'_{++}(\pi)=0$. Therefore, the helicity of the electron has
to be flipped, but {\it its spin remains conserved}. At longer
wave length $\lambda=10R$, $\sigma'_{+-}$ gains more weight and
the total differential cross-section
$\sigma'_+=\sigma'_{++}+\sigma'_{+-}$ becomes more isotropic.

\begin{figure}
\includegraphics[width=3.in]{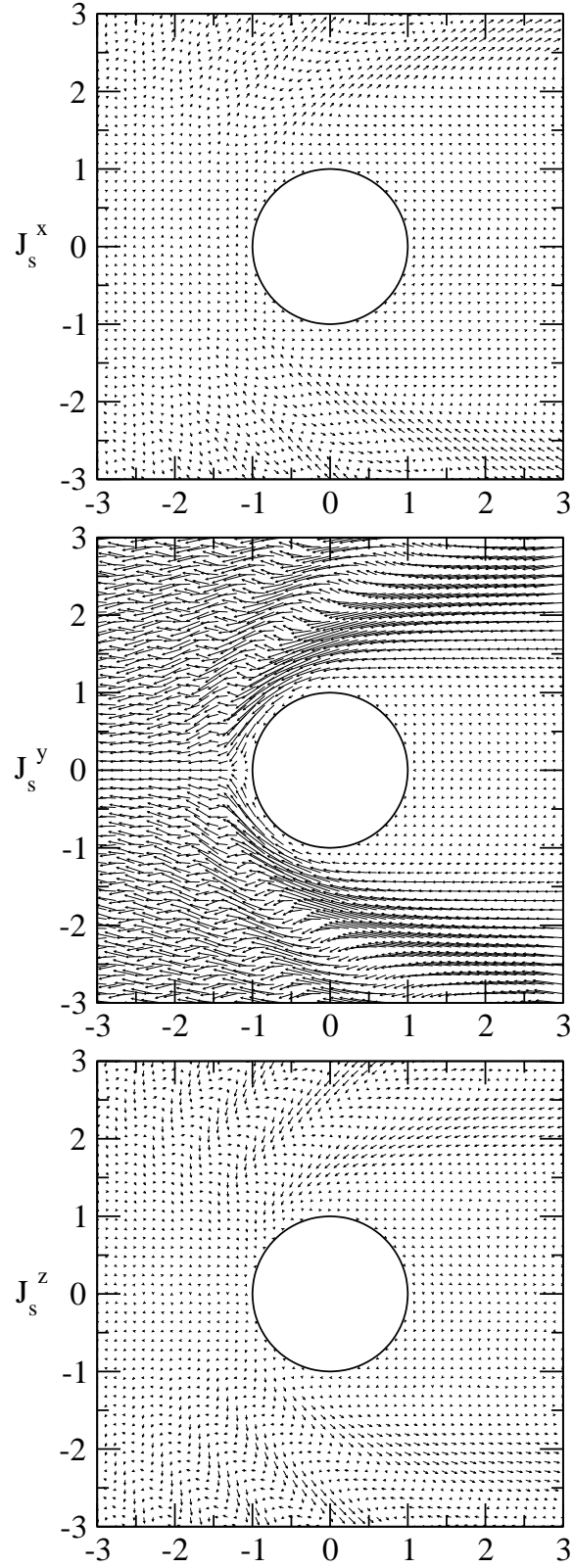}
\caption{Distributions of spin current density ${\vec j}^x_s$,
${\vec j}^y_s$, and ${\vec j}^z_s$. The incident wave has a
well-defined helicity $\eta=+$ with $\lambda_{\rm in}=R$.
[$\alpha=(1/10)\hbar^2k/2m$]} \label{fig6}
\end{figure}

The distributions of spin current density are shown in
Fig.~4.\cite{old} A prominent feature in the figures is the
overall trend for the $\vec{j}_s^y(\vec{r})$ vectors to point to
the left. This counter-intuitive behavior is simply related to the
fact that the spin current density is equal to the product of
velocity and spin, where the velocity is in the positive
$x$-direction and the spin points to the {\it minus} $y$-direction
for $\Psi_{{\rm in},+}$, therefore the scattering $\vec{j}_s^y$
vectors generically point to the left. On the contrary, for
$\Psi_{{\rm in},-}$ with spin points to the {\it positive}
$y$-direction, the direction of the flow will be reversed. This
also explains why the magnitudes of the spin currents
$\vec{j}_s^{x,z}$ are small in most of the regions, since the
original incident current has no $S_x$ and $S_z$ components.

In Fig.~4, it can be seen that $\vec{j}_s^z$ oscillates both in
amplitude and direction between the curved stripes. An incident
wave with opposite helicity would reverse such a flow. Therefore,
part of these flows are cancelled if the incident current is not
polarized. However, local spin current that oscillates in space
still exist, similar to the case of the spin density in Fig.~1.
All of the local spin currents $\vec{j}_s^{x,y,z}$ would vanish
for unpolarized incident electrons once the Rashba coupling is
turned off. However, if $\alpha$ is nonzero but the disk is
removed (i.e., free space), then there exist spin currents
$\vec{j}_s^x=\alpha/2\hat{x}$, $\vec{j}_s^y=-\alpha/2\hat{y}$, and
$\vec{j}_s^z=\vec{0}$, which is the background spin current
cautioned by Rashba.\cite{rashba} It reflects the unsatisfying
current status on a proper definition of the spin current.

Our system with unilateral current flow seems to be the same as
the 2DEG driven by an electric field,\cite{sinova} but there is no
global transverse spin current $\vec{j}_s^z$, no matter the disk
is removed or not. This does not contradict the result of the
proposed intrinsic spin Hall effect in a clean 2DEG.\cite{sinova}
The incident charge current flowing to the right can be understood
as originating from the slightly unbalanced electrochemical
potentials on the two leads far away. In our case, all the
electrons are moving along the direction of the potential
gradient, instead of moving at all directions on the Fermi surface
in Sinova {\it et al}'s paper.\cite{sinova} Consequently, no spin
Hall effect is expected if one follows similar semiclassical
analysis in Ref.~\onlinecite{sinova}.

\section{Conclusion}

In summary, the influence of the spin-orbit coupling on the
scattering of 2D electrons from a hard disk is studied. Such a
simple setup offers us a good opportunity to investigate the
properties of spin and spin current in details. We focus our
attention on the near-field regime, where the scattered wave is
comparable to the incident wave, and appreciable out-of-plane
spins can be found. This work offers us a clear understanding of
the microscopic dynamics around the mesoscopic disk, and could
serve as a basis for future works considering a threaded magnetic
flux in the disk, or a hybrid device involving a disk as a
component.

\acknowledgements The authors acknowledge the support from the
National Science Council in Taiwan. M.C.C. thanks M.F. Yang for
helpful discussions.

\end{document}